\documentclass[conference]{IEEEtran}
\IEEEoverridecommandlockouts
\usepackage{amsmath,amssymb,amsthm,amsfonts,amsopn,cite,mathrsfs,todonotes}
\usepackage[latin1]{inputenc}
\usepackage[T1]{fontenc}
\usepackage{ifpdf}
\usepackage[english]{babel}
\usepackage{subfig} 
\usepackage[ruled]{algorithm2e}
\usepackage{dsfont}

\usepackage{color,url}
\usepackage{hyperref}
\newcounter{tempEquationCounter}
\newcounter{thisEquationNumber}
\newenvironment{floatEq}
{\setcounter{thisEquationNumber}{\value{equation}}\addtocounter{equation}{1}
\begin{figure*}[tb]
\hrulefill\vspace*{4pt}
\normalsize\setcounter{tempEquationCounter}{\value{equation}}
\setcounter{equation}{\value{thisEquationNumber}}
}
{\setcounter{equation}{\value{tempEquationCounter}}
\hrulefill\vspace*{-1em}
\end{figure*}
}

\theoremstyle{plain}

\theoremstyle{definition}
\theoremstyle{remark}
\theoremstyle{remark}
\newtheorem*{rem*}{Remark}

\newcommand{\bd}{\mathbf}   

\newcommand{\LtR}{\mathbf L^2(\mathbb{R})}

\newcommand{\RR}{\mathbb R}
\newcommand{\ZZ}{\mathbb Z}

\def\sgn{\mathop{\operatorname{sgn}}}

\DeclareFontFamily{U}{mathx}{\hyphenchar\font45}
\DeclareFontShape{U}{mathx}{m}{n}{
      <5> <6> <7> <8> <9> <10>
      <10.95> <12> <14.4> <17.28> <20.74> <24.88>
      mathx10
      }{}
\DeclareSymbolFont{mathx}{U}{mathx}{m}{n}
\DeclareFontSubstitution{U}{mathx}{m}{n}
\DeclareMathAccent{\widecheck}{0}{mathx}{"71}
\DeclareMathAccent{\wideparen}{0}{mathx}{"75}

\newcommand{\Glg}{{\mathcal G(\gamma,g)}}
\newcommand{\GlTg}{{\mathcal G(\gamma,\bd T g)}}
\newcommand{\GlDg}{{\mathcal G(\gamma,g')}}
\newcommand{\GlTtg}{{\mathcal G(\gamma,\bd T^2 g)}}
\newcommand{\GlTDg}{{\mathcal G(\gamma,\bd T g')}}
\newcommand{\Glgn}{{\mathcal G(\gamma,g_0)}}

\newcommand{\GlTtgn}{{\mathcal G(\gamma,\bd T^2 g_0)}}

\newcommand{\Parx}{\frac{\partial}{\partial x}}

\newcommand{\Parxi}{\frac{\partial}{\partial \xi}}

\newcommand{\phasesyn}{\phi_\text{s}}

\begin{document}

\author{\IEEEauthorblockN{Zden\v ek Pr\r u\v sa, Nicki Holighaus}
\IEEEauthorblockA{Acoustics Research Institute, Austrian Academy of Sciences\\ 
Wohllebengasse 12--14, A-1040 Vienna, Austria\\ E-Mail: 
\{zdenek.prusa,nicki.holighaus\}@oeaw.ac.at}}
 
 \title{Non-iterative Filter Bank Phase (Re)Construction
 \thanks{This work was supported in 
part by the Austrian Science Fund (FWF) START-project FLAME (``Frames and Linear 
Operators for Acoustical Modeling and Parameter Estimation''; Y 551-N13) and 
MERLIN (I 3067-N30).\\
\emph{This is the Author's Accepted Manuscript version of work presented at EUSIPCO17. It is licensed under the terms of the \href{https://creativecommons.org/licenses/by/4.0/}{Creative Commons Attribution 4.0 International License}, which
permits unrestricted use, distribution, and reproduction in any medium, provided the original author and source are credited. The published version is available at:} \url{https://ieeexplore.ieee.org/abstract/document/8081342}}}
\maketitle
 
 \begin{abstract}
   Signal reconstruction from magnitude-only measurements
  presents a long-standing problem in signal processing. 
In this contribution, we propose a phase 
(re)construction method for filter banks with uniform decimation and controlled 
frequency variation.
The suggested procedure extends the 
recently introduced phase-gradient heap integration 
and relies 
on a phase-magnitude relationship for filter bank coefficients obtained 
from Gaussian filters. Admissible filter banks are modeled as the discretization 
of certain generalized translation-invariant systems, 
for which 
we derive the phase-magnitude relationship explicitly. The 
implementation for discrete signals is 
described and the performance of the algorithm is evaluated on a range of real 
and synthetic signals.
 \end{abstract} 
  
  \section{Introduction}
  
  In this contribution, we suggest a direct method for the construction 
  of time-frequency phase information from magnitude-only measurements with respect
  to a collection of analysis filters. 
  In Fourier-based signal analysis, phase information
  is crucial for signal reconstruction from filter bank (FB) coefficients. 
  %
  Two variants of the phase reconstruction problem are most prominent: 
  (a) Due to limitations in the measurement/analysis process, only magnitude measurements can be obtained or the phase is involuntarily lost in some processing step. 
  (b) In many processing applications, the phase of the analysis coefficients before processing is known. However, after coefficient
   modification, the known phase is often invalid and has to be adjusted.
  While the first instance is common in optics and medical imaging, where phase retrieval has been an active problem 
  for several decades~\cite{gesa72}, the second instance is arguably more important in audio signal processing. It 
  arises in applications such as source separation and denoising \cite{guse10},
  time-stretching/pitch shifting \cite{lado99}, speech synthesis 
  \cite{rif17} and missing data inpainting \cite{smbima11}, to name a few. 
  
  In order to reconstruct a signal from representation coefficients, it is necessary for the underlying representation to be invertible. For linear systems, invertibility is essentially equivalent to the frame property. Moreover, it has been shown that for a generic phase retrieval algorithm to have any hope of providing reliable solutions, a certain overcompleteness is strictly necessary~\cite{bacaed05}. For such redundant, invertible linear systems, a number of iterative phase retrieval schemes have been proposed, the most important of
  which is the Griffin-Lim algorithm (GLA)~\cite{griflim84}. In particular, the recent fast GLA
  (fGLA)~\cite{pebaso13} provides good results with reasonable computational performance. Generally, all iterative phase reconstruction algorithms require a significant number of rather costly iterations, see also~\cite{desomada15} for an alternative to fGLA . 
  For the particular case of the short-time Fourier transform (STFT), specialized methods have been
  presented~\cite{leroux10,zhbewy07,be15,ltfatnote040}. Here, we introduce an extension of phase gradient heap integration (PGHI), see~\cite{ltfatnote040}, where a more exhaustive overview and comparison of previous phase reconstruction schemes is given. PGHI uses the phase-magnitude relationship of the STFT with a Gaussian window~\cite{po79} to compute the phase gradient from the magnitude coefficients and generate a phase estimate by integration. 
   
  We derive a generalization of the essential equations provided in~\cite{ltfatnote040}, valid for
  certain generalized translation-invariant (GTI) systems~\cite{jale14}. Although we are not able to exactly determine the phase gradient solely from known information, our evaluation shows that the resulting approximation achieves excellent results.
  
  \textbf{Notation:}
  In this manuscript, we consider continuous or discrete signals of finite energy, i.e. $s\in\LtR$ or $s\in\ell^2(\ZZ)$. By $\bd T_x$, $\bd M_\xi$ and $\bd D_\gamma$ we denote the translation, modulation and dilation operators given by 
   $\bd T_x s = s(\cdot-x)$, $\bd M_\xi s = e^{2\pi i \xi (\cdot)} s$,\text{ and } $\bd
   D_\gamma s = \gamma^{-1/2} s(\cdot/\gamma)$ 
  and their analogue on $\ell^2(\ZZ)$. Without
  subscript, $\bd T$ denotes the time-weighting operator $\bd T s = (\cdot)s$.
   
  \section{GTI systems with controlled frequency 
variation and the derived filter banks}   
A generalized 
translation-invariant (GTI) system on $\LtR$ is a collection of functions 
$\{g_i\}_{i\in I}\subset \LtR$, for some index set $I$, 
together with all their translations on the real line, i.e. 
    $
     \left\{\bd 
T_x g_i\right\}_{x\in\RR,i\in I}.
    $ 
    Here, we only consider $I=\RR$, identified with 
frequency, and $g_\xi := \bd M_\xi \bd D_{\gamma(\xi)} g$, where $g\in\LtR$ is 
the \emph{prototype function} and $\gamma:\RR\rightarrow \RR^+$ is a continuous 
function of the frequency variable $\xi$ determining the 
\emph{frequency-bandwidth relationship}. We define
    \begin{equation}\label{eq:defOfStructGTI}
      \mathcal G(\gamma,g) := \left\{g_{x,\xi}\right\}_{x,\xi\in\RR}, \text{ 
where } g_{x,\xi}:= \bd T_x \bd M_\xi \bd D_{\gamma(\xi)} g.
    \end{equation}
    The analysis coefficients of a function $s$ 
    with respect to $\mathcal 
G(\gamma,g)$ are defined through the inner products  
    \begin{equation}\label{eq:GTIcoeffs}
      \begin{split}
      c_s(x,\xi):= V_{\mathcal G(\gamma,g)} s(x,\xi) := \langle 
s,g_{x,\xi}\rangle
       = s\ast \overline{g_\xi(-\cdot)}(x),
      \end{split}
    \end{equation}
    for all $x,\xi\in\RR$. The final equality shows that $c_s(\cdot,\xi)$ is a 
filtering of $s$ with the filter $\overline{g_\xi(-\cdot)}$.
    The complex-valued function $V_{\mathcal G(\gamma,g)} s$ can be described by 
its magnitude and phase as  
    \begin{equation}
     V_{\mathcal G(\gamma,g)} s(x,\xi) = M_{\mathcal G(\gamma,g)} 
s(x,\xi)e^{2\pi i\phi^s_{\mathcal G(\gamma,g)}(x,\xi)},
    \end{equation}
    for all $x,\xi\in\RR$, where $M_{\mathcal G(\gamma,g)} s := |V_{\mathcal G(\gamma,g)} s|$ is the 
magnitude and $\phi^s_{\mathcal G(\gamma,g)}$ is the phase of $V_{\mathcal 
G(\gamma,g)} s$.
    
      Let $\{g_k\}_{k\in\ZZ}\subset\LtR$ be a collection of functions and $a\in\RR^+$ 
a \emph{decimation factor}. The system $\{g_{n,k}\}_{n,k\in\ZZ}$ with $g_{n,k} = \bd T_{na}g_k$ is a \emph{filter bank 
(FB)}. The analysis coefficients of $s$
with respect to $\{ g_{n,k} 
\}_{n,k\in\ZZ}$ are 
      \[
       c_s[n,k] = \langle s,g_{n,k}\rangle.
      \]      
      A FB is said to form a frame, if there are constants $0 < A\leq B 
< \infty$, 
      such that $A\|s\|_2^2\leq \|c_s\|^2_2 \leq B\|s\|_2^2, \text{ for all } 
s\in\LtR$. The frame property guarantees the stable invertibility of the 
coefficient mapping by means of a dual frame 
$\{\widetilde{g_{n,k}}\}_{n,k\in\ZZ}$, i.e. 
      \begin{equation}\label{eq:frsyn}
        s = \sum_{n,k} c_s[n,k]\widetilde{g_{n,k}}, \text{ for all } s\in\LtR.
      \end{equation}
      For FBs with uniform decimation, various efficient methods exist for 
computing the dual frame or at least the synthesis operation \eqref{eq:frsyn}, see~\cite{bofehl98,gr93,necciari17}.      
      Clearly, if $\xi:\ZZ\rightarrow \RR$ is an increasing function, the filter 
bank
      \begin{equation}\label{eq:defOfStructFB}
       \mathcal G(\gamma\circ \xi,g,a) := \{ \bd T_{na} \bd M_{\xi(k)} \bd 
D_{\gamma(\xi(k))} g\}_{n,k\in\ZZ}
      \end{equation}
      is a sampling of $\mathcal G(\gamma,g)$ and 
      \[
       \begin{split}
       c_s[n,k] & = V_{\mathcal G(\gamma,g)} s(na,\xi(k))\\
       & = M_{\mathcal G(\gamma,g)} s(na,\xi(k))e^{2\pi i\phi^s_{\mathcal 
G(\gamma,g)}(na,\xi(k))}.
       \end{split}
      \]
  
  \section{Phase-magnitude relationships for Gaussian GTI systems}
   Assume that $g\in\mathcal C^1$ and 
$\gamma\in\mathcal C^1$. A straightforward calculation using
\begin{equation}\label{eq:VGlogTimeDeriv}
   \begin{split}
   \Parx \log\left(V_\Glg s (x,\xi)\right) & = \Parx \log\left(M_\Glg 
s(x,\xi)\right)\\
   & \hspace{12pt} + i\Parx \phi^s_\Glg(x,\xi),
   \end{split}
   \end{equation}
   and analogous for $\Parxi \log\left(V_\Glg s\right)$, show that the equalities provided
   in \eqref{eq:derivExpr} hold for all $g\in\LtR\cap\mathcal C^1(\RR)$. The derivation steps here are analogous to \cite{auchfl12}. Now, if we set $g = g_0 := 2^{1/4}e^{-\pi(\cdot)^2}$, then the equality $g_0' = -2\pi\bd T g_0$ yields \eqref{eq:gaussMagPhase} and if $\gamma$ is constant, we obtain the phase-magnitude relationship for the STFT~\cite{po79}. 
   \begin{floatEq}
  \begin{equation}\label{eq:derivExpr}
   \begin{split}
     \Parx \phi^s_\Glg(x,\xi) & = 
       \xi-\frac{1}{2\pi\gamma(\xi)}\text{Im}\left(\frac{V_\GlDg s(x,\xi)}{V_\Glg 
s(x,\xi)}\right)\\
     \Parxi \phi^s_\Glg(x,\xi) & = 
-\frac{\gamma'(\xi)}{2\pi\gamma(\xi)}\text{Im}\left(\frac{V_\GlTDg s(x,\xi)}{V_\Glg 
s(x,\xi)}\right)\ - \gamma(\xi)\text{Re}\left(\frac{V_\GlTg s(x,\xi)}{V_\Glg 
s(x,\xi)}\right)\\
     \Parx \log(M_\Glg s)(x,\xi) & = 
-\gamma(\xi)^{-1}\text{Re}\left(\frac{V_\GlDg s(x,\xi)}{V_\Glg 
s(x,\xi)}\right)\\
     \Parxi \log(M_\Glg s)(x,\xi) & = -\frac{\gamma'(\xi)}{2\gamma(\xi)} - 
\frac{\gamma'(\xi)}{\gamma(\xi)}\text{Re}\left(\frac{V_\GlTDg s(x,\xi)}{V_\Glg 
s(x,\xi)}\right) + 2\pi\gamma(\xi)\text{Im}\left(\frac{V_\GlTg s(x,\xi)}{V_\Glg 
s(x,\xi)}\right).
   \end{split}
  \end{equation}
  \hrulefill\vspace*{4pt}
  \begin{equation}\label{eq:gaussMagPhase}
   \begin{split}
     \Parx \phi^s_\Glgn(x,\xi) & = \xi+\frac{\Parxi \log(M_\Glgn 
s)(x,\xi)}{4\pi^2 
\gamma(\xi)^2}+\frac{\gamma'(\xi)}{4\pi\gamma(\xi)^3}-\frac{\gamma'(\xi)}{
2\pi\gamma(\xi)^3}\text{Re}\left(\frac{V_\GlTtgn s(x,\xi)}{V_\Glgn 
s(x,\xi)}\right)\\
       \Parxi \phi^s_\Glgn(x,\xi) & = - \gamma(\xi)^2 \Parx \log(M_\Glgn 
s)(x,\xi) + \frac{\gamma'(\xi)}{4\pi^2\gamma(\xi)}\text{Im}\left(\frac{V_\GlTtgn 
s(x,\xi)}{V_\Glgn s(x,\xi)}\right).
   \end{split}
  \end{equation}
  \hrulefill\vspace*{4pt}
  \begin{equation}\label{eq:IntTimeForward}
     \widetilde{\phi}^s_\Glgn[n\pm 1,k] := \widetilde{\phi}^s_\Glgn[n,k] \pm
      a\left(\frac{\Delta^{\phi,x,s}_\Glgn[n\pm 1,k]+\Delta^{\phi,x,s}_\Glgn[n,k]}{2}\right) 
   \end{equation}
   \begin{equation}\label{eq:IntFreqForward}
     \widetilde{\phi}^s_\Glgn[n,k\pm 1] := \widetilde{\phi}^s_\Glgn[n,k] \pm |\xi(k\pm
       1)-\xi(k)|\left(\frac{\Delta^{\phi,\xi,s}_\Glgn[n,k\pm 1]+\Delta^{\phi,\xi,s}_\Glgn[n,k]}{2}\right) 
   \end{equation}
   \setcounter{tempEquationCounter}{\value{equation}}
    \end{floatEq}    
  
  \section{Application to filter bank phase (re)construction}
  Given a FB $\mathcal G(\gamma\circ \xi,g_0,a)$ as per 
\eqref{eq:defOfStructFB}, the results of the previous section can be used to 
obtain a phase estimate $\widetilde{\phi}^s_\Glg$ from the magnitude 
measurements $|c_s[n,k]| = M_{\mathcal G(\gamma,g_0)} s(na,\xi(k))$. 
To that end, PGHI~\cite{ltfatnote040} is adapted to 
cope with the more general filter bank structure. 
  Before a phase estimate can be constructed, we have to compute an estimate of 
the phase derivative from the given magnitude. Assuming that only $|c_s|$, 
$\xi(\cdot)$ and $\gamma(\xi(\cdot))$ are known, we have 
  \begin{equation*}\begin{split}
    \lefteqn{\Parx \phi^s_\Glgn(na,\xi(k)) \approx 
      \Delta^{\phi,x,s}_\Glgn[n,k]}\\
    & \hspace{14pt} := 2\pi\xi(k) +\frac{\Delta_k 
      \left(\gamma\circ\xi\right)(k)}{2\gamma(\xi(k))^3}
      +\frac{\Delta_k 
      \left(\log(|c_s|)\right)(n,k)}{2\pi\gamma(\xi(k))^{2}},\\ 
    \lefteqn{\Parxi \phi^s_\Glgn(na,\xi(k)) \approx 
      \Delta^{\phi,\xi,s}_\Glgn[n,k]}\\
    & \hspace{14pt} := - 2\pi\gamma(\xi(k))^{2} \Delta_n 
      \left(\log(|c_s|)\right)(n,k),
   \end{split}\end{equation*}
   where $\Delta_n$ and $\Delta_k$ are discrete differentiation schemes. 
   Since the sampling step in time is uniform and equals the decimation
   factor $a$, we use simple centered differences for $\Delta_n$, i.e. 
   \begin{equation}
     \Delta_n(c)(n) : = \frac{c(n+1)-c(n-1)}{2a}.
   \end{equation}
   The sampling step in frequency is variable, depending on $\xi(\cdot)$. Hence, 
   weighted centered differences are used:
   \begin{equation}
     \Delta_k(c)(k) : = \frac{c(k+1)-c(k)}{2(\xi(k+1)-\xi(k))}-\frac{c(k)-c(k-1)}{2(\xi(k)-\xi(k-1))}.
   \end{equation}   
   Note that we omit the terms depending on 
   $
   V_\GlTtg s
   $. Although this introduces 
additional inaccuracies, our results have shown that the contribution of those 
terms is minor and their omission has little adverse effect.
   
   The integration of $(\Delta^{\phi,x,s}_\Glgn[n,k],\Delta^{\phi,\xi,s}_\Glgn[n,k])$ to obtain 
   an estimate for $\phi^s_\Glgn(na,\xi(k))$ is performed through $1$D trapezoidal 
   quadrature. Integration in time direction is again straightforward, see \eqref{eq:IntTimeForward}, while integration in frequency direction takes the channel distance into account \eqref{eq:IntFreqForward}. The purpose of the heap integration algorithm, described in the next section, is the initialization of the integration and the adaptive selection of the integration path, i.e. when to use \eqref{eq:IntTimeForward} or \eqref{eq:IntFreqForward}. 
   
  \section{Implementation and Analysis}  
  In practice, we work with sampled signals and digital filters in $\ell^2(\ZZ)$. The signal $s$ and the prototype filter $g$ are assumed to be samples of smooth and localized functions, such that the procedure described above still provides a valid estimate of the phase of $c_s$. Moreover, we only have to consider a limited number of frequency channels $k\in\underline{K}:=\{0,\ldots,K-1\}$ and, if $s$ is finitely supported, time positions $n\in\underline{N}:=\{0,\ldots,N-1\}$, for which to compute $\widetilde{\phi}^s_\Glgn(na,\xi(k))$. Algorithm \ref{alg:fbpghi} (FBPGHI), a modified version of PGHI, is used to compute the phase estimate $\widetilde{\phi}^s_\Glgn$. If $N$ is very large, or $s$ has infinite support, RTPGHI~\cite{ltfatnote043} can be similarly adapted.

  \begin{algorithm}[h!t]
    \LinesNumbered
    \caption{Phase Gradient Heap Integration - FBPGHI}
    \label{alg:fbpghi}
    \KwIn{Magnitude $|c_s|$ of FB coefficients, estimates $\Delta^{\phi,x,s}_\Glg$
    and $\Delta^{\phi,\xi,s}_\Glg$ of the time and frequency phase derivative, relative tolerance $\mathit{tol}$.}
    \KwOut{Phase estimate $\widetilde{\phi}^s_\Glgn$.}    
    $\mathit{abstol} \leftarrow \mathit{tol}\cdot \max\left( c_s[n,k] \right)$\;
    Create set $\mathcal{I}=\left\{(n,k)\in \underline{K}\times\underline{N} : c_s[n,k]>\mathit{abstol}
    \right\}$\;\label{alg:setI}
    Assign random values to $\phasesyn(n,k)$ for $k\notin \mathcal{I}$\;
    Construct a self-sorting max \emph{heap} \cite{wi64} for $(n,k)$ pairs\;\label{alg:line}   
    \While{$\mathcal{I}$ is not $\emptyset$  }{
        \If{\emph{heap is empty}}{
                Move $(k_m,n_m)=\mathop{\operatorname{arg~max}}\limits_{(n,k)\in\mathcal I} \left(\left|c_s[n,k]\right|\right)$ from $\mathcal I$ into the \emph{heap}\;
                $\widetilde{\phi}^s_\Glgn(k_m,n_m) \leftarrow 0$\;                
        }
        \While{\emph{heap} is not empty}{
            $(n,k) \leftarrow$ remove the top of the
            \emph{heap}\;\label{alg:heapremove}
            
            \If{$(n,k+1)\in\mathcal I$}{
                Apply Eq. \eqref{eq:IntTimeForward}(+)\;
                Move $(n,k+1)$ from  $\mathcal{I}$ into the \emph{heap}\;
            }
            \If{$(n,k-1)\in\mathcal I$}{
                Apply Eq. \eqref{eq:IntTimeForward}(-)\;
                Move $(n,k-1)$ from  $\mathcal{I}$ into the \emph{heap}\;
            } 

            \If{$(k+1,n)\in\mathcal I$}{
                Apply Eq. \eqref{eq:IntFreqForward}(+)\;
                Move $(k+1,n)$ from  $\mathcal{I}$ into the \emph{heap}\;                
            }
            \If{$(k-1,n)\in\mathcal I$}{
                Apply Eq. \eqref{eq:IntFreqForward}(-)\;
                Move $(k-1,n)$ from  $\mathcal{I}$ into the \emph{heap}\;
            }            
        }
    }  
    \bigskip
\end{algorithm}  
  The algorithm introduces several possible sources of inaccuracy. In addition to the errors present in PGHI, we
  (a) approximate $\gamma'(\xi(k))$ by a weighted centered difference only involving $\gamma(\xi(k)),\gamma(\xi(k\pm1))$, where available.
  (b) disregard the real or imaginary part of $\frac{\gamma'(\xi)V_\GlTtg s(x,\xi)}{\gamma(\xi)V_\Glg s(x,\xi)}$ in \eqref{eq:gaussMagPhase}, 
  since there is no straightforward way to obtain them from known information.  
  Therefore, the accuracy of the algorithm rests on $\gamma$ not to vary too quickly. In this case, the derivative of $\gamma$ is approximated well by the finite difference scheme and moreover, the factor $\gamma'(\xi)/\gamma(\xi)$ is expected to be small, such that the missing term has 
  little influence on the result.
  
  \section{Evaluation}
  
\begin{figure}[t]
    \includegraphics[height=0.2\textwidth,trim=0cm 0cm 0cm 0.1cm, clip]{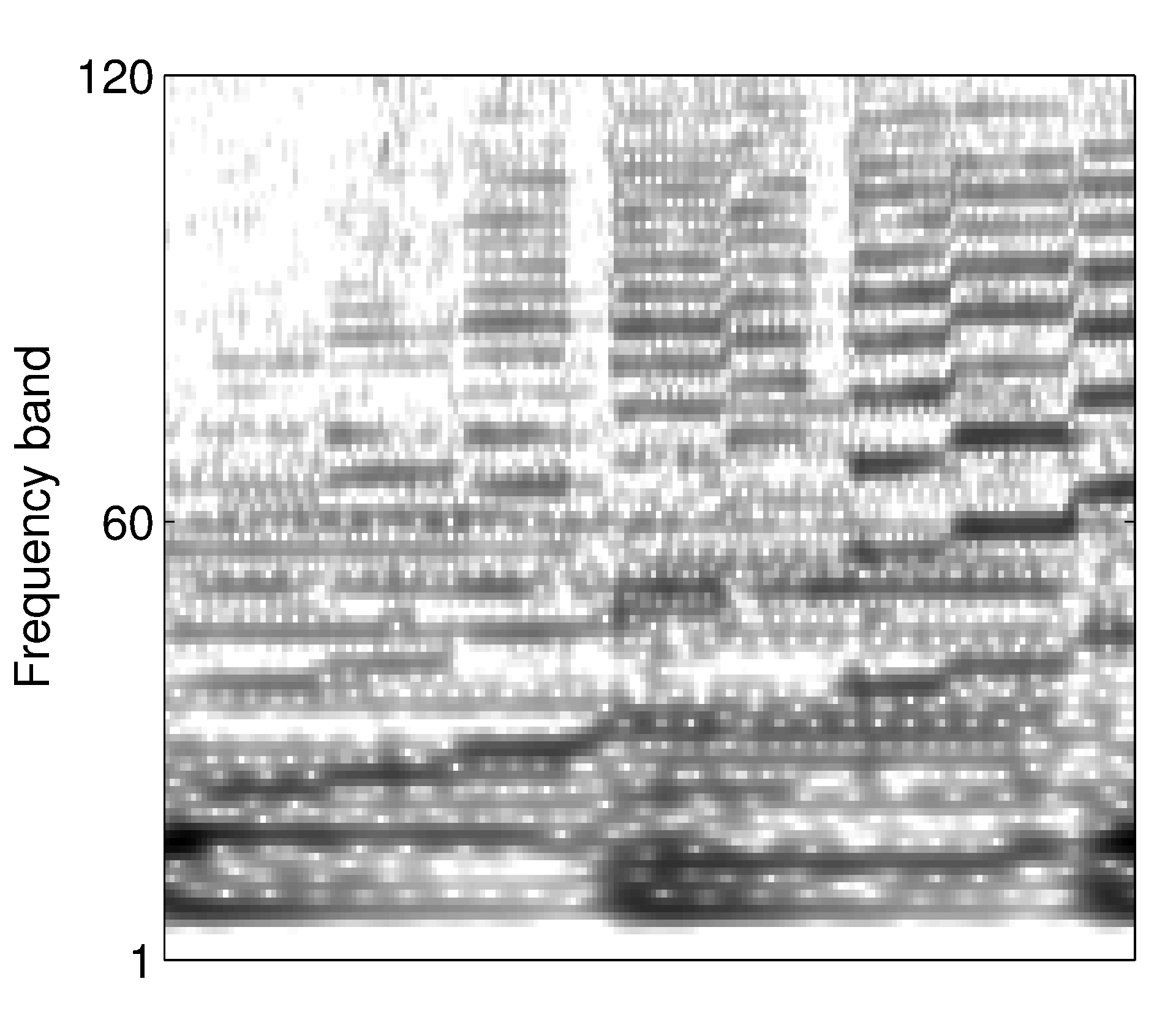}\hspace{0.6em}\includegraphics[height=0.2\textwidth,trim=0cm 0cm 0cm 0cm, clip]{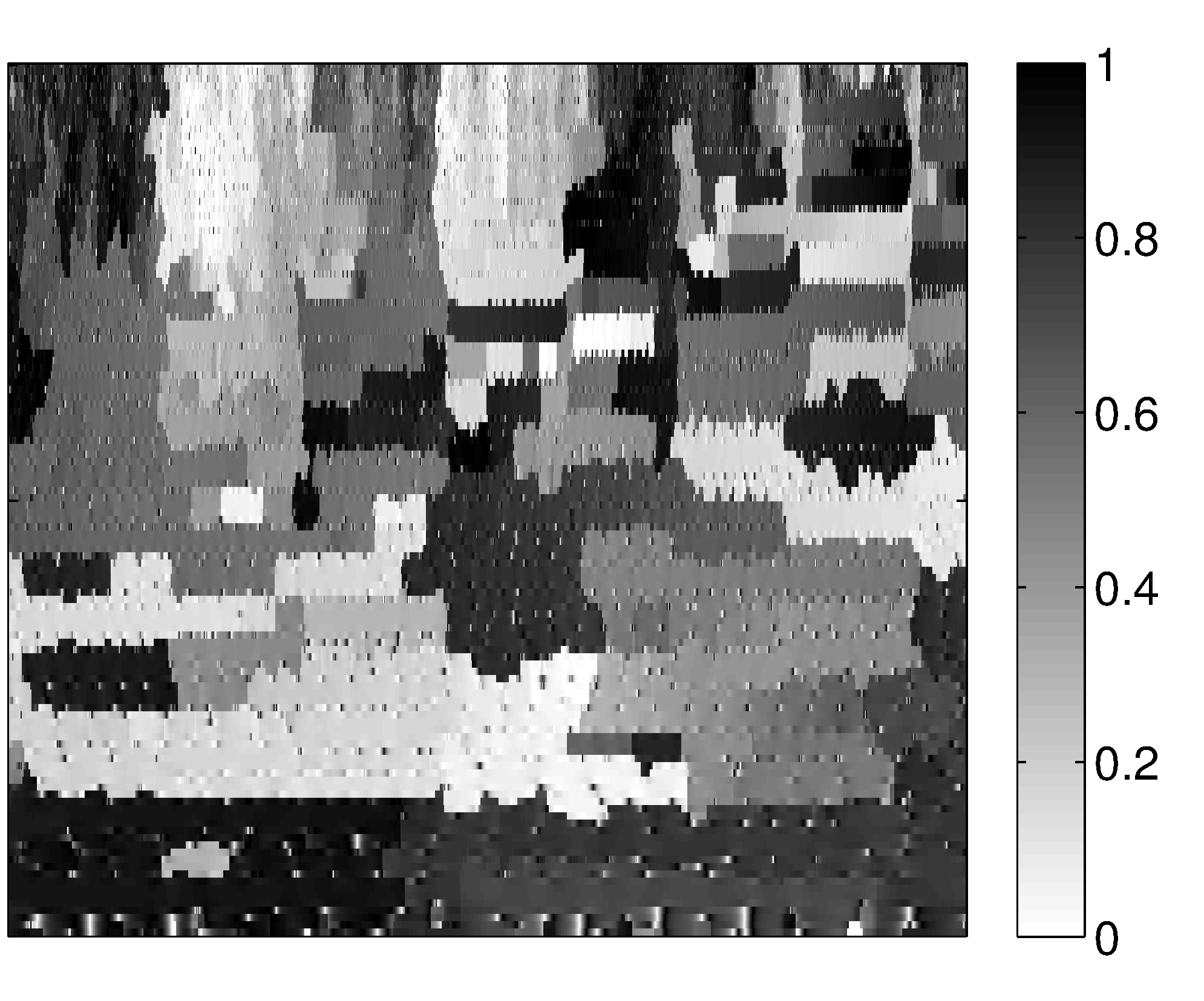}
   \caption{Filter bank spectrogram (l) and difference between original and FBPGHI-restored phase (r) for an excerpt of $s_4$. The displayed phase difference is the difference between the phase angles in radians, divided by $\pi$.}\label{fig:phasedifference}
  \end{figure}

  To demonstrate the performance of the proposed algorithm, we applied it to a number of
  real and synthetic audio signals, using several different filter bank configurations.
  Moreover, we compare our results to the results provided by the established, iterative
  \emph{fast Griffin-Lim}  algorithm (fGLA). Phase reconstruction
  algorithms are usually not expected to recover the original phase exactly and,
  typically, the reconstruction quality cannot be easily judged by simply comparing the waveforms of
  the original and reconstructed signals. In~\cite{griflim84}, Griffin and Lim have 
  proposed to use the spectral difference 
  $
    E_{\text{spec}}(s,\tilde{s}) = 20\log_{10}\left(\frac{\||c_s| - |c_{\tilde{s}}|\|_2}{\|c_s\|_2}\right),
  $
  to measure the distortion of the phase-restored signal $\tilde{s}$. 
  Despite some flaws, 
  $E_{\text{spec}}$ usually provides a useful indicator of the restoration quality. 
  Figure \ref{fig:phasedifference}(r) shows a typical example of the phase difference between the original representation phase and the phase obtained with FBPGHI. See \cite{ltfatnote040} for more details.  
  
  \textbf{Evaluation setup:}
  For the evaluation, we selected seven signals, sampled at $\xi_s=44.1$~kHz each:
  \begin{itemize}
   \item $s_1[l] = \sum_{k=0}^7 \sin(220\pi \cdot 2^k l/\xi_s)$, $l\in\underline{\xi_s}$.
   \item $s_2[l] = \sum_{k=0}^3 \sin(220\pi \cdot 2^{2k} l/\xi_s) + \sum_{k=1}^8 \delta_{5000k}[l] + \text{echirp}_1[l]+\text{echirp}_2[l]$, $l\in\underline{\xi_s}$, where $\text{echirp}_1$ and $\text{echirp}_2$ are real-valued, constant amplitude chirps with exponential frequency modulation and center frequency increasing from $500$~Hz to $15$~kHz, resp. decreasing from $18$~kHz to $3$~kHz.
   \item $s_3$ is $\xi_s$ samples of white noise.
   \item $s_4$ to $s_7$ comprise 4 second excerpts of a jazz recording (brass and percussion), signal number $54$ from the SQAM database~\cite{sqam} (male German speech), Ophelia's Song by Musetta and a classical Indian melody (both female singing voice).
  \end{itemize}
  For each signal, we applied the proposed algorithm and fGLA in five different FB configurations. All FB choices have in common that the filter center frequencies and bandwidths are chosen with respect to a given frequency scale, i.e. we have $bins$ filters per scale unit with a bandwidth of $bw$ scale units. 
  The tested configurations are as follows:
  \begin{enumerate}
   \item Adapted to the ERB scale~\cite{Glasberg:1990a}, $bins = 1$ and $bw = 2$, for the full frequency range, see also~\cite{necciari17}.
   \item Adapted to the ERB scale, but $bins = 4$, $bw = 1/2$.
   \item Adapted to the scale $10\log(\xi)$, i.e. constant-Q, $bins = 4$ and $bw=1/2$, with minimum frequency $30$~Hz and maximum frequency $\xi_s/2$, see~\cite{dogrhove13}.
   \item Adapted to the scale $\sgn(\xi)\left((1+|\xi/4|)^{1/2}-1\right)$, with parameters identical to the constant-Q FB.
   \item Adapted to the scale $8\sgn(\xi)\left((1+|\xi|)^{1/4}-1\right)$, with parameters identical to the constant-Q FB.
  \end{enumerate}
  The latter $2$ scales have no particular perceptual relevance and were chosen merely for demonstration purposes.  
  For now, our method only considers uniform decimation by $a$. 
  The chosen decimation factors and resulting redundancies are shown in Table \ref{tab:1}.
    
  \textbf{Quantitative evaluation:}    
  Table~\ref{tab:2} 
  lists the spectral difference in dB of the solution provided by the proposed algorithm for all combinations of signals and filter banks. It can be seen that, despite considerably different redundancies, the algorithm performs similar for all considered FBs in terms of spectral difference. The possible exception to this rule is the ERB-scale FB(1) with only $1$ filter per ERB, which performs worse in almost all cases. Also of note are the large values of $E_{\text{spec}}$ for the noise $s_3$, which is consistent with the evaluations in \cite{ltfatnote040}.
  \begin{table}[t]
   \begin{center}
   \begin{tabular}{||c||c|c|c|c|c||}
   \hline
   FB  & (1) & (2) & (3) & (4) & (5)\\
   \hline
   $a$ & 8 & 36 & 20 & 73 & 33 \\ 
   $R$ & 10.75 & 9.44 & 26.40 & 8.00 & 21.58\\  
   \hline    
   \end{tabular}
   \end{center}
   \caption{Decimation factor and redundancy for the considered filter banks.}\label{tab:1}
   \vspace*{4pt}
   \begin{center}
   {\setlength{\tabcolsep}{0.33em}
   \begin{tabular}{||c||c|c|c|c|c|c|c||}
   \hline
   FB  & $s_1$ & $s_2$ & $s_3$ & $s_4$ & $s_5$ & $s_6$ & $s_7$\\
   \hline
   (1) & $-25.65$ & $-24.87$ & $-12.25$ & $-20.22$ & $-26.61$ & $-28.42$ & $-26.38$\\
   (2) & $-32.62$ & $-28.76$ & $-12.89$ & $-23.24$ & $-26.96$ & $-31.44$ & $-28.70$\\
   (3) & $-34.75$ & $-29.21$ & $-14.30$ & $-23.96$ & $-27.65$ & $-30.28$ & $-28.49$\\
   (4) & $-34.52$ & $-30.76$ & $-14.39$ & $-23.08$ & $-25.70$ & $-32.59$ & $-29.28$\\
   (5) & $-35.72$ & $-31.32$ & $-15.93$ & $-23.41$ & $-28.03$ & $-33.15$ & $-29.93$\\
   \hline    
   \end{tabular}
   }
   \end{center}
   \caption{FBPGHI - Spectral difference values $E_{\text{spec}}$ in dB.
   }\label{tab:2}
   \vspace{-1em}
   \end{table}
   
\begin{figure}[ht]
      \centering
    \includegraphics[width=0.40\textwidth,trim=0cm 2cm 0cm 1.2cm, clip]{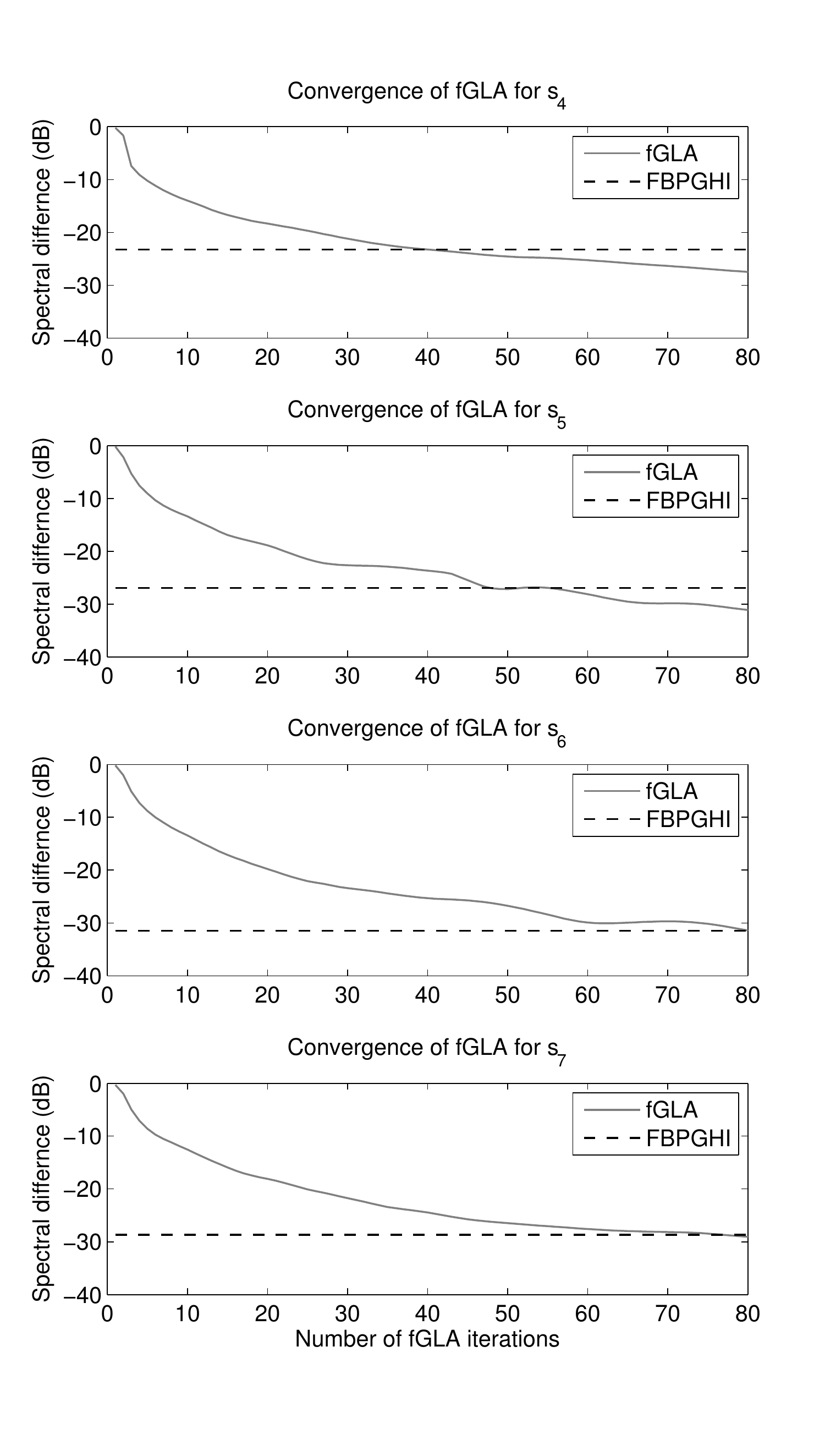}
    \caption{Comparison of fGLA convergence and FBPGHI. Results were obtained with FB(2).}\label{fig:fbpghiVsFglim}
    \vspace{-1em}
  \end{figure}

  \textbf{Comparison with iterative methods:}  
  In \cite{ltfatnote040,pebaso13}, it was shown that, for Gabor
  transforms, fGLA performs comparably or better than other iterative schemes in terms of
  spectral difference. Since we have no reason to assume that the situation changes in the
  filter bank setting, we consider fGLA as reference algorithm. In Figure \ref{fig:fbpghiVsFglim}, we provide some examples as to how FBPGHI compares to
  fGLA iterations for the signals $s_4$ to $s_7$ and FB(2). Between $30$ and $80$ fGLA steps 
  are necessary to achieve the same $E_{\text{spec}}$ as FBPGHI and any meaningful improvement
  requires a large number of additional fGLA steps\footnote[1]{For the full set of comparisons, audio examples and extended experiments, please refer to the supplementary material at
  \url{http://ltfat.github.io/notes/051/}.}. Note that every step of fGLA requires $1$
  application each of FB synthesis and analysis. Therefore, every iteration has
  considerable computational cost, while FBPGHI is very efficient,  
  see also \cite{ltfatnote040} for more details. In the same contribution, it was shown that 
  the initialization of fGLA with PGHI provided a significant quality boost over both methods. We expect the same for FBPGHI.
  
  \textbf{Perceptual performance:}
  Informal listening of the signals restored by the proposed FBPGHI algorithm or $80$ fGLA iterations, 
  for signals $s_1$ through $s_7$ and FBs (1) to (5) led to the conclusion that
  both methods performed comparably and without significant artifacts on all signals\footnotemark[1].
  No clear performance gap between the algorithms could be detected, with the exception of FB(1) which produced clearly audible artifacts for both FBPGHI and fGLA,
  albeit on different signals. We attribute these artifacts to the poor frequency resolution of FB(1).   
  
  \section{Conclusions and Outlook}
  We have provided an extension of the recent PGHI algorithm
  for phase reconstruction to filter banks with controlled frequency variation and uniform
  decimation. Experiments have shown that the algorithm performs competitively in terms of an established objective error measure and also perceptually.
  
  A significant drawback of the proposed method is the required redundancy, in particular
  for filter banks with highly varying filter bandwidths, see Table \ref{tab:1} FBs (3),(5). 
  Therefore, a logical next step
  will be the combination of the heap integration method with nonuniform decimation. Such
  a scheme will enable the selection of an appropriate sampling step for each frequency
  channel. Therefore, it can be expected that redundancy is significantly reduced
  without meaningful impact to the restoration quality. However, the adaptation of the
  heap integration method to a truly nonuniform sampling grid requires significant work. 
  
  In \cite{ltfatnote050}, the authors propose an improved phase vocoder based on 
  PGHI for time-stretching and pitch-shifting. 
  A similar application of FBPGHI is conceivable and might possibly further improve the quality of the achieved effect. Future theoretic work could be concerned with finding an appropriate estimate including the neglected terms in \eqref{eq:gaussMagPhase} as well as estimates for the error if the used filters differ from the Gaussian. Preliminary results\footnotemark[1] have shown promising results for Blackman filters.
  
  \section*{Acknowledgment}
This work was supported by the Austrian Science Fund (FWF): Y~551--N13 and I~3067--N30.
  
  \bibliographystyle{IEEEtran}
  \bibliography{addbib}  
\end{document}